\documentclass[10pt]{article}

\usepackage[margin=1in]{geometry}
\usepackage{microtype}
\usepackage{lmodern}
\usepackage[T1]{fontenc}
\usepackage{csquotes}
\usepackage[english]{babel}
\usepackage{xcolor}
\usepackage{graphicx}
\usepackage{booktabs}
\usepackage{amsmath, amssymb}
\usepackage{siunitx}
\usepackage{enumitem}
\usepackage{caption}
\usepackage{subcaption}
\usepackage{float}
\usepackage{hyperref}
\usepackage[noabbrev,nameinlink,capitalise]{cleveref}
\usepackage{longtable}
\usepackage{booktabs}

\hypersetup{
  colorlinks=true,
  linkcolor=blue!50!black,
  citecolor=blue!50!black,
  urlcolor=blue!70!black
}

\usepackage[backend=biber,style=authoryear,maxbibnames=9,
  natbib=true,uniquename=false,giveninits=true]{biblatex}
\graphicspath{{figures/}}

\title{The New Quant: A Survey of Large Language Models in Financial Prediction and Trading}
\author{Weilong Fu\thanks{Columbia University, USA. \texttt{wf2232@columbia.edu}}}
\date{ }

\begin{document}
\maketitle

\begin{abstract}
Large language models are reshaping quantitative investing by turning unstructured financial information into evidence-grounded signals and executable decisions. This survey synthesizes research with a focus on equity return prediction and trading, consolidating insights from domain surveys and more than fifty primary studies. We propose a task-centered taxonomy that spans sentiment and event extraction, numerical and economic reasoning, multimodal understanding, retrieval-augmented generation, time series prompting, and agentic systems that coordinate tools for research, backtesting, and execution. We review empirical evidence for predictability, highlight design patterns that improve faithfulness such as retrieval first prompting and tool-verified numerics, and explain how signals feed portfolio construction under exposure, turnover, and capacity controls. We assess benchmarks and datasets for prediction and trading and outline desiderata-for time safe and economically meaningful evaluation that reports costs, latency, and capacity. We analyze challenges that matter in production, including temporal leakage, hallucination, data coverage and structure, deployment economics, interpretability, governance, and safety. The survey closes with recommendations for standardizing evaluation, building auditable pipelines, and advancing multilingual and cross-market research so that language-driven systems deliver robust and risk-controlled performance in practice.
\end{abstract}



\section{Introduction}
\label{sec:intro}

Large Language Models enable a shift from feature-centric text mining to end-to-end decision systems in markets. We refer to this emerging paradigm as the new quant, by which we mean investment processes where language models read and reason over heterogeneous disclosures, generate auditable hypotheses, interact with external tools and data, and translate textual understanding into risk-controlled positions. This survey concentrates on the pipeline components that matter most for investment outcomes, namely financial prediction with an emphasis on equity return prediction and trading with portfolio construction. We systematize advances from 2023 to 2025 through this lens \parencite{Zhao2024Revolutionizing,Lee2024FinLLMsSurvey,Nie2024FinancialLLMsSurvey,Liu2024SurveyFinancialAI,Kong2024LLMInvestment,Xu2024GenAIEnterprise}.

Transformer pretraining and instruction tuning produced general purpose models with non-trivial reasoning and tool use \parencite{Devlin2019BERT,Brown2020GPT3,Achiam2023GPT4}. Domain-specific financial language models and open weight ecosystems make finance-grade adaptation feasible under privacy, governance, and cost constraints \parencite{Wu2023BloombergGPT,Scao2023BLOOM,Touvron2023LLaMA,Touvron2023LLaMA2,MetaAI2024Llama3}. Efficient tuning through low-rank adaptation, quantization-sensitive fine-tuning, and one-bit optimization lowers the barrier to controlled deployment in trading environments \parencite{Hu2021LoRA,Dettmers2024QLoRA,Ma2024OneBitLLMs}. In parallel, task-level capabilities have matured across sentiment, information extraction and knowledge graphs, numerical question answering, long document understanding, multi-modal analysis, and agentic decision support. These capabilities feed, constrain, and explain predictive signals and trade decisions \parencite{Kong2024LLMInvestment,Nie2024FinancialLLMsSurvey,Zhao2024Revolutionizing}.

Evidence already suggests that model derived views on news, filings, earnings calls, and policy communications can predict returns in certain settings, although evaluation practice often falls short of trading standards. Leakage control, stress testing, market microstructure realism, and cost or capacity reporting remain inconsistent. Governance and interpretability requirements, such as evidence-based rationales, audit logs, and a clear separation between signal generation and portfolio allocation, are likewise unevenly addressed.

Our contributions are fourfold. First, we frame the design space for the development of new quantitative and review models relevant to finance together with the efficiency techniques that make financial language models practical \parencite{Devlin2019BERT,Brown2020GPT3,Achiam2023GPT4,Wu2023BloombergGPT,Hu2021LoRA,Dettmers2024QLoRA,Ma2024OneBitLLMs}. Second, we offer a task taxonomy centered on prediction and trading that clarifies how the upstream natural language processing components feed tradable signals \parencite{Nie2024FinancialLLMsSurvey,Zhao2024Revolutionizing,Kong2024LLMInvestment}. Third, we synthesize the literature on return prediction in Section~\ref{sec:return-pred} and on trading with portfolio construction in Section~\ref{sec:trading-portfolio}. We cover interpretable financial language models, retrieval augmented pipelines, time series aware prompting, and multi-agent trading systems. Fourth, we consolidate benchmarks and datasets in Section~\ref{sec:benchmarks} and articulate challenges in Section~\ref{sec:challenges} that involve temporal leakage, faithfulness, evaluation realism, cost and latency, and governance. The audience includes researchers who build financial language models for tradable use cases, quantitative practitioners who evaluate language model signals, and leaders who design audit-ready deployment strategies \parencite{Xu2024GenAIEnterprise,Kong2024LLMInvestment}.

\section{Foundations for Prediction and Trading with FinLLMs}
\label{sec:lm-dev}

\subsection{From transformers to tool using language models}
The transformer replaced recurrent networks with attention and enabled scalable pretraining on large corpora \parencite{Vaswani2017Attention}. Decoder only GPT models demonstrated emergent in context and few shot abilities \parencite{Radford2018GPT1,Radford2019GPT2,Brown2020GPT3}. Encoder only models such as BERT and RoBERTa delivered state of the art text understanding for classification and span extraction \parencite{Devlin2019BERT,Liu2019RoBERTa}. The GPT\mbox{\,}4 technical report and instruction tuning advances, as exemplified by FLAN, established language models as general purpose controllers with meaningful reasoning and tool use \parencite{Achiam2023GPT4,Wei2021FLAN}.

\subsection{Open models and efficient adaptation for finance}
Open releases including BLOOM and the LLaMA families catalyzed a flourishing ecosystem that supports controlled adaptation and on premise deployment \parencite{Scao2023BLOOM,Touvron2023LLaMA,Touvron2023LLaMA2,MetaAI2024Llama3}. Additional families such as Qwen, Baichuan, and InternLM expand the menu of base models \parencite{Bai2023Qwen,Baichuan2023,InternLM2024}. Efficient training and post training alignment through low rank adaptation, quantization aware finetuning, and one bit optimization enable domain tuned models with modest compute budgets, which aligns with privacy and reproducibility constraints that are common in trading environments \parencite{Hu2021LoRA,Dettmers2024QLoRA,Ma2024OneBitLLMs}.

\subsection{Financial PLMs and FinLLMs}
Before instruction following language models, financial applications built on encoder style pretrained language models such as the FinBERT line and showed early domain transfer benefits \parencite{Araci2019FinBERT,Yang2020FinBERTComm,Liu2021FinBERTIJCAI}. Recent finance specific language models include BloombergGPT \parencite{Wu2023BloombergGPT}, PIXIU \parencite{Xie2023PIXIU}, FinGPT \parencite{Yang2023FinGPT,Liu2024FinGPTHPC}, InvestLM \parencite{Yang2023InvestLM}, Instruct FinGPT \parencite{Zhang2023InstructFinGPT}, DISC FinLLM \parencite{Chen2023DISCFinLLM}, and CFGPT \parencite{Li2023CFGPT}. Language and market specific adaptations such as SilverSight and FinVisGPT tailor models to regional corpora and workflows \parencite{Zhou2024SilverSight,Wang2023FinVisGPT}. FinTral reports a multimodal family with performance at the reported level of general purpose models \parencite{Bhatia2024FinTral}. New compact bases such as Mistral\mbox{\,}7B have also been adopted in finance settings \parencite{Jiang2023Mistral7B}.

\subsection{Implications for prediction and trading}
For return prediction, decoder models support rationale generation and in context composition over news, filings, and macro text, while encoder models remain strong for narrow sentiment or extraction tasks that feed signals. Hybrid systems that combine retrieval augmented language models, language driven graph or sequence models, and mixture of experts pipelines appear frequently, often coupled with faithfulness checks and backtesting aware evaluation \parencite{Kong2024LLMInvestment,Nie2024FinancialLLMsSurvey}. For trading, agentic frameworks with tool use, memory, and role specialization begin to structure research, critique, and execution under constraints, which motivates new benchmarks and simulation protocols that we discuss in later sections.

\section{Task taxonomy for financial prediction and trading}
\label{sec:tasks}

This section codifies a taxonomy that maps language model capabilities to finance workflows and clarifies how prediction and trading depend on upstream natural language processing. We group tasks by the primary function they serve in production pipelines, while noting that deployed systems often combine several capabilities in a single workflow.

\subsection{Sentiment and opinion as signal inputs}
The objective is to infer polarity, stance, and intensity from heterogeneous sources such as news, social media, earnings calls, and analyst notes, and to transform these assessments into features for event studies, return prediction, or risk monitoring. Domain tuned encoders in the FinBERT line demonstrate strong transfer on finance texts \parencite{Araci2019FinBERT,Yang2020FinBERTComm,Liu2021FinBERTIJCAI}. Instruction tuned language models can score sentiment and produce justifications and they sometimes outperform classical lexicon baselines on complex material \parencite{LopezLira2023ChatGPTStocks,Steinert2023Microblogging,Luo2024FinancialSentimentLLMs}. Work on FOMC minutes and ECB press conferences indicates that policy tone can be quantified and linked to market responses \parencite{Gossi2023FinBERTFOMC,Kanelis2024ECBIndicator}. Classic resources remain useful as baselines and diagnostic tools \parencite{Loughran2011LM,Pennebaker2001LIWC,Stone1966GeneralInquirer,Mishev2020EvalSentimentFinance,Tan2023SentimentSurvey,Bordoloi2023SentimentSurvey}. In a trading context sentiment features must be timestamped, free of look ahead leakage, and aligned to realistic rebalancing schedules.

\subsection{Information extraction and knowledge graphs for point in time signals}
Information extraction converts unstructured documents into structured entities, relations, and events that can feed screens, factor engines, and retrieval modules. Datasets such as FiNER, FinRED, and REFinD support supervised training and enable point in time knowledge curation \parencite{Hillebrand2022KPIBERT,Shah2023FiNERDataset,Sharma2022FinRED,Kaur2023REFinD}. Large language models can assist IE through prompting or lightweight finetuning for named entity recognition and relation extraction \parencite{Covas2023GPTNER,Rajpoot2023GPTFinRE}. Event detection and relation modeling in Chinese and English demonstrate cross market applicability \parencite{Tian2019ChineseEventRE,Wan2023CFERE}. As institutions deploy knowledge graphs for research, LLMs act as controllers and generators that populate and query the graphs while surveys outline integration patterns and governance requirements \parencite{WeaverBird2023,Jiang2023KGSurvey,Pan2023LLMKG,Pan2024TKDERoadmap,Li2023FinDKG,Liang2024AlignLLMGraphDB,Deloitte2020KGPoV}. For prediction and trading these components provide upstream signal generation and retrieval that improves evidence quality.

\subsection{Numerical question answering and reasoning for thesis validation}
These systems execute multi step reasoning over tables, text, and formulas in filings, earnings calls, and macro releases and they answer questions while computing key performance indicators. Benchmarks such as FinQA, FinanceBench, BizBench, DocMathEval, and EconLogicQA probe numerical correctness, long document understanding, and economics logic that underlies valuation and surprise based signals \parencite{Chen2021FinQA,Islam2023FinanceBench,Koncel2023BizBench,Zhao2023DocMathEval,Quan2024EconLogicQA}. Retrieval augmented generation with layout aware encoders and tool calls improves verifiability and accuracy and verification or constrained decoding reduces hallucination risk \parencite{Phogat2023ZeroShotQA,Srivastava2024MathReasoning,Arun2023NumericalReasoning,Wang2024DocLLM}. In production pipelines these systems are most valuable when they produce intermediate calculations and citations that can be audited before the trade decision.

\subsection{Summarization and document understanding for evidence condensation}
Abstractive and extractive hybrids and instruction tuned models can compress long financial narratives such as ten K filings, management discussion and analysis, and earnings calls, which accelerates research and supports hypothesis generation \parencite{ElHaj2019MultiLing,LaQuatra2020EndToEndSumm,Mukherjee2022ECTSum,Abdaljalil2021SummarizationFinance,Zmandar2021LongFinSumm}. Retrieval aware chunking and long document architectures reduce information loss and yield more stable summaries \parencite{Yepes2024RAGChunk,Beltagy2020Longformer}. For trading use the outputs should be materiality aware and time stamped and they should reference evidence spans that analysts can verify.

\subsection{Multimodal cues for predictive signals}
Beyond text, systems fuse audio from calls, visuals such as charts, or structured time series to inform predictive modeling. Datasets and models for multimodal analysis of earnings calls and policy communication supply prosodic and visual cues \parencite{Li2020MAEC,Mathur2022MONOPOLY}. FinVisGPT addresses chart reading and explanation, and language model informed graph or sequence models use textual context to guide stock movement prediction \parencite{Wang2023FinVisGPT,Chen2023ChatGPTGNN,Wimmer2023VisionLanguageMarkets}. RiskLabs illustrates multi source fusion for risk prediction \parencite{RiskLabs2024}. For trading deployment multimodal models must meet latency constraints and must ensure that any audio or visual evidence is available at decision time.

\subsection{Agentic workflows for trading and execution}
Agentic frameworks operationalize language models as decision support agents with memory, tools, and role specialization. TradingGPT introduces layered memory and distinct analyst characters and later systems expand tool use, multi agent debate, and evaluation in leakage controlled simulations \parencite{Li2023TradingGPT,Zhang2024FinAgent,Yu2024FinMem,Wang2024QuantAgent,Wang2023AlphaGPT,Yuan2024AlphaGPT2}. Surveys of LLM based agents, computational experiments, memory mechanisms, and trust or safety provide design guidance and highlight open problems such as planning reliability and tool misuse \parencite{Xi2023LLMAgentsSurvey,Guo2024MultiAgentsSurvey,Ma2024CompExperimentsAgents,Zhang2024MemorySurvey,Hua2024TrustAgent}. In practice these agents should separate research from order routing and they should log prompts, retrievals, and tool calls for audit.

\subsection{Governance functions that constrain trading systems}
Financial institutions explore LLMs for auditing support, contradiction detection, and regulatory interpretation \parencite{Berger2023AuditLLM,Deusser2023Contradictions,CaoFeinstein2024RegInterpret,ChoiKim2024TaxAudits}. These capabilities do not execute trades and they do shape acceptable model behavior, guardrails, and evidence requirements for production systems. Trading oriented deployments benefit from explicit policies that bind language to timestamped evidence and that restrict action when verification fails.

\begin{table}[t]
\centering
\caption{Mapping of tasks to trading relevance}
\label{tab:tasks_to_trading}
\begin{tabular}{@{}p{3.0cm}p{4.0cm}p{3.2cm}p{4.6cm}@{}}
\toprule
Task & Representative artifacts & Typical outputs & Contribution to trading \\
\midrule
Sentiment and opinion & FinBERT line, instruction tuned LLM scorers & Polarity, stance, justifications & Event features, risk monitoring, regime filters \\
\midrule
Information extraction and knowledge graphs & FiNER, FinRED, REFinD, FinDKG, WeaverBird & Entities, relations, events, KG triples & Point in time factors, high precision retrieval, constraint checks \\
\midrule
Numerical QA and reasoning & FinQA, FinanceBench, DocMathEval, EconLogicQA & Computed KPIs, verified answers, reasoning chains & Thesis validation, surprise based signals, audit trails \\
\midrule
Summarization and document understanding & ECTSum, MultiLing FNS, Longformer, RAG chunking & Condensed briefs with citations & Faster research, explanation for signals and trades \\
\midrule
Multimodal analysis & MAEC, MONOPOLY, FinVisGPT & Prosody features, chart readings, fused embeddings & Additional cues for selection or sizing under latency limits \\
\midrule
Agentic workflows & TradingGPT, FinAgent, FinMem, QuantAgent & Tool calls, debate traces, memory states & Orchestration of research, verification, and execution \\
\midrule
Governance and compliance & Audit and regulation tools & Contradiction flags, policy checks & Guardrails that shape allowable actions and documentation \\
\bottomrule
\end{tabular}
\end{table}

\section{LLMs for return prediction}
\label{sec:return-pred}

This section surveys how language models produce equity return signals and how those signals can be translated into investable decisions. We organize methods by evidence channels and modeling patterns and then discuss evaluation practice and practical guidance. We focus on studies from 2023 to 2025 with direct implications for trading.

\subsection{Problem formulation and evidence channels}
The objective is to map text and related modalities to expected returns at horizons that range from intraday to monthly. Common channels include news and social media, corporate disclosures and earnings calls, and policy communications and macro releases. Work using general purpose language models reports that zero or few shot prompts on event text can be predictive in some settings \parencite{LopezLira2023ChatGPTStocks,Steinert2023Microblogging}. Domain specific corpora such as earnings call transcripts provide richer cues in the narrative and question and answer segments and several studies evaluate models on this setting \parencite{Cook2023LocalLLMs}. Central bank statements and press conferences also encode information that matters for assets that are sensitive to interest rates and risk appetite \parencite{Gossi2023FinBERTFOMC,Kanelis2024ECBIndicator}.

\subsection{Modeling patterns for text to return signals}
\paragraph{Zero and few shot scoring with general models}
General purpose models can be prompted to classify event direction or to assign a return score together with a rationale. Early finance studies report out of sample predictability from such scores on news and social media \parencite{LopezLira2023ChatGPTStocks,Steinert2023Microblogging}. Practical systems often add calibration layers, confidence filtering, and symbol mapping before portfolio construction.

\paragraph{Domain and instruction tuned FinLLMs}
Instruction tuning on finance instructions and curated corpora improves robustness and reduces prompt brittleness. Representative models include InvestLM, Instruct FinGPT, FinGPT and its high performance computing variants, and FinLlama \parencite{Yang2023InvestLM,Zhang2023InstructFinGPT,Yang2023FinGPT,Liu2024FinGPTHPC,Konstantinidis2024FinLlama}. Value aligned or preference tuned variants have also been explored \parencite{Yu2024GreedLlama}. These models typically strengthen sentiment and event classification and they can generate explanations that are easier to audit.

\paragraph{Retrieval augmented modeling and knowledge grounded signals}
Retrieval augmented generation reduces hallucination risk and improves faithfulness by binding predictions to timestamped evidence. Financial pipelines add retrieval aware chunking and layout features for long documents \parencite{Yepes2024RAGChunk,Wang2024DocLLM}. Knowledge graphs and retrieval over company specific graphs further structure the evidence and stabilize factor construction \parencite{WeaverBird2023,Li2023FinDKG}. RAG enhanced sentiment has been shown to improve downstream accuracy on finance tasks \parencite{Zhang2023RAGFinSent}.

\paragraph{LLM guided structured models}
Language models can supply features or supervisory signals to graph and sequence models that are optimized for price movement prediction. Examples include graph neural networks whose edges or node priors are informed by language model judgments and vision language systems that read price charts \parencite{Chen2023ChatGPTGNN,Wimmer2023VisionLanguageMarkets}. Risk oriented work fuses model derived signals with other sources to predict adverse events \parencite{RiskLabs2024}.

\paragraph{Time series aware prompting and forecasting}
Several studies examine how to connect language models to time series forecasting. Approaches include reprogramming prompts for temporal inputs, using language models as zero shot forecasters, and using specialized long horizon transformer architectures alongside language models for reasoning and explanation \parencite{Jin2023TimeLLM,Gruver2024ZeroShotTS,Zhou2022FEDformer,Nie2022Times64,Wen2022TransformersTS,Liang2024FMTSurvey}. Although these papers are not always finance specific, the techniques inform how to condition signals on regime and horizon.

\paragraph{Interpretable designs and auditability}
Production systems require transparent rationales and clear provenance for each predicted effect. Interpretable pipelines generate structured explanations and highlight evidence spans and they expose ablation or counterfactual checks. Recent work proposes interpretable stock movement modeling with finance specific rationale templates and self reflective explanations \parencite{Tong2024Ploutos,Koa2024SelfReflect}. These designs help risk teams to understand when and why a signal should be trusted.

\subsection{Evaluation protocols that meet trading standards}
Return prediction requires time safe evaluation. We recommend rolling walk forward splits with document availability enforced at decision time and with an embargo to prevent label leakage from post event commentary. Report signal quality with correlation and calibration and report economics with returns that include explicit commission and spread assumptions, Sharpe and drawdown, turnover and capacity, and sensitivity to universe and rebalancing frequency. Given growing evidence of look ahead issues in pretrained models, evaluation should check for time machine effects using dated corpora and explicit filters \parencite{Sarkar2024Lookahead,Drinkall2024TimeMachineGPT}. Baselines should include naive and factor models and trend benchmarks to avoid overstating the incremental value of language signals \parencite{Jiang2023PriceTrends}.

\subsection{What works when and practical guidance}
Language signals often add value around identifiable events and narrative changes and they can complement price based factors during regime shifts. News and social media sentiment tends to matter at shorter horizons when coverage is fast and dense. Earnings call analysis matters at announcement and in the following days when management tone and detail resolve uncertainty. Policy communication sentiment is most relevant for rate sensitive sectors and for broad risk appetite proxies. In all cases the portfolio should separate signal generation from allocation and risk and it should include materiality filters, confidence gating, and exposure controls.

\begin{center}
\begin{longtable}{@{}p{4.0cm}p{3.2cm}p{7.5cm}@{}}
\caption{Representative papers for equity return prediction with one sentence summaries}
\label{tab:return_pred_papers_long} \\
\toprule
Paper & Setting or channel & One sentence summary \\
\midrule
\endfirsthead

\multicolumn{3}{l}{\textit{Table \thetable\ continued}}\\
\toprule
Paper & Setting or channel & One sentence summary \\
\midrule
\endhead

\midrule
\multicolumn{3}{r}{\textit{Continued on next page}}\\
\endfoot

\bottomrule
\endlastfoot

\textcite{LopezLira2023ChatGPTStocks} & News and filings & Zero and few shot scores from a general model predict cross sectional returns in several universes with controls for headline leakage. \\
\midrule
\textcite{Steinert2023Microblogging} & Social media & Microblog sentiment from a large model correlates with next day stock moves and improves on lexicon baselines. \\
\midrule
\textcite{Cook2023LocalLLMs} & Earnings calls & Locally hosted language models score call tone and deliver signals that survive controls for known factors. \\
\midrule
\textcite{Gossi2023FinBERTFOMC} & Policy minutes & FinBERT tuned for policy text extracts sentiment from FOMC minutes that aligns with market responses. \\
\midrule
\textcite{Kanelis2024ECBIndicator} & Press conferences & A sentiment indicator from ECB statements explains euro area asset movements and complements macro variables. \\
\midrule
\textcite{Yang2023InvestLM} & Finance tuned LLM & Instruction tuned InvestLM improves investment specific judgments and produces auditable rationales. \\
\midrule
\textcite{Zhang2023InstructFinGPT} & Finance tuned LLM & Instruct FinGPT strengthens finance sentiment and can act as a robust scoring component in pipelines. \\
\midrule
\textcite{Yang2023FinGPT,Liu2024FinGPTHPC} & Open finance LLM & FinGPT provides open models and recipes that enable cost aware domain adaptation for finance tasks. \\
\midrule
\textcite{Konstantinidis2024FinLlama} & Sentiment for trading & FinLlama demonstrates instruction tuned scoring for trading oriented sentiment classification. \\
\midrule
\textcite{Yu2024GreedLlama} & Preference tuned LLM & GreedLlama studies value alignment for financial reasoning and highlights the effect on moral or risk trade offs. \\
\midrule
\textcite{Yepes2024RAGChunk} & Retrieval and chunking & Retrieval aware chunking improves long document question answering for filings and earnings analysis. \\
\midrule
\textcite{Wang2024DocLLM} & Layout aware modeling & A layout aware generator improves numerical reasoning over tables and reduces errors in KPI extraction. \\
\midrule
\textcite{WeaverBird2023} & Knowledge grounded RAG & A system that couples language models with a knowledge base and search engine improves decision support quality. \\
\midrule
\textcite{Li2023FinDKG} & Dynamic knowledge graphs & A dynamic finance knowledge graph supports point in time retrieval for research and signal construction. \\
\midrule
\textcite{Chen2023ChatGPTGNN} & Text guided GNN & A graph neural network informed by language model judgments improves stock movement prediction. \\
\midrule
\textcite{Wimmer2023VisionLanguageMarkets} & Vision language & A vision language approach uses chart images to detect granular market changes that relate to returns. \\
\midrule
\textcite{RiskLabs2024} & Multi source risk & A multi source pipeline with a language model integrates diverse data to predict financial risk events. \\
\midrule
\textcite{Jin2023TimeLLM} & Time series prompting & A reprogramming approach adapts language models to time series forecasting and yields competitive accuracy. \\
\midrule
\textcite{Gruver2024ZeroShotTS} & Zero shot forecasting & Large language models used as zero shot forecasters provide reasonable baselines for several temporal datasets. \\
\midrule
\textcite{Zhou2022FEDformer} & Long horizon TS & A frequency enhanced transformer delivers strong long horizon forecasting and can complement language signals. \\
\midrule
\textcite{Nie2022Times64} & Tokenization for TS & A tokenization approach converts time series into compact sequences that are well suited to transformers. \\
\midrule
\textcite{Tong2024Ploutos} & Interpretable stock movement & An interpretable finance specific model produces rationales that link text spans to predicted movement. \\
\midrule
\textcite{Koa2024SelfReflect} & Self reflective explanations & A method that uses self reflection yields explainable stock predictions with improved plausibility of rationales. \\
\midrule
\textcite{Jiang2023PriceTrends} & Baseline for trends & A comprehensive study of trend models supplies strong baselines that are useful when measuring incremental value. \\
\end{longtable}
\end{center}

\section{LLM assisted trading systems and portfolio construction}
\label{sec:trading-portfolio}

This section analyzes how language models support trading decisions from idea generation to execution and how they interact with portfolio construction. We organize the discussion around the life cycle of a trade and we emphasize designs that produce auditable, time safe, and economically meaningful outcomes.

\subsection{From assisted research to executable strategies}
Agentic systems transform language models into research assistants that read disclosures, propose hypotheses, and coordinate tools such as retrieval, calculators, and backtesters. TradingGPT introduces layered memory and distinct analyst roles that debate and refine theses before handing off to tools \parencite{Li2023TradingGPT}. FinAgent expands the toolkit to include multimodal inputs and broker like actions under a tool governance layer \parencite{Zhang2024FinAgent}. FinMem focuses on memory design that stabilizes multi day workflows and preserves analyst intent during iteration \parencite{Yu2024FinMem}. QuantAgent explores self improvement loops that critique prompts and strategies and that then retest within a controlled simulator \parencite{Wang2024QuantAgent}. Alpha GPT and its successor Alpha GPT 2.0 formalize analyst in the loop alpha discovery with critique, ranking, and evaluation gates to reduce overfitting \parencite{Wang2023AlphaGPT,Yuan2024AlphaGPT2}. Together these systems show how assisted research can evolve into executable strategies while keeping human oversight in the loop.

\subsection{Prompting and language to strategy}
Several studies convert natural language descriptions into screen definitions, factor recipes, or backtest scripts. Work on code generation for trading strategies indicates that language models can scaffold usable code with human review and unit tests \parencite{Noguer2024CodeGenTrading}. Conversational research tools support exploratory analysis and rapid what if checks for fundamental and event driven theses \parencite{Yue2023GPTQuant}. Effective practice includes canonicalizing prompts into machine readable templates, validating data access permissions, and compiling prompts and code into immutable artifacts that can be audited later.

\subsection{Retrieval verified analysis loops}
Hallucination and numerical brittleness motivate retrieval verified workflows. Retrieval aware chunking and layout aware modeling improve KPI extraction from filings and reduce reasoning errors in long documents \parencite{Yepes2024RAGChunk,Wang2024DocLLM}. Systems that couple a language model with a curated knowledge base and a search engine demonstrate higher faithfulness for decision support \parencite{WeaverBird2023}. In trading contexts the loop proceeds as propose, retrieve, verify, and only then simulate or trade. Each step produces traces with timestamps and evidence spans to support review by risk and compliance.

\subsection{From signals to orders and execution}
Language models that score text still require a conversion to orders and an execution policy that respects market microstructure. A practical pattern separates signal generation from order placement and routing. Execution quality depends on latency, slippage, queue priority, and the balance between limit and market orders. Recent work on generative modeling for limit order book message flow offers realistic simulators for policy testing \parencite{Nagy2023GenAILOB}. Decision systems should log order intents, parameter choices, and realized costs to enable attribution and continuous improvement.

\subsection{Portfolio construction with language model support}
Portfolio construction benefits from language models in two ways. First, LLM derived signals enter a classical optimizer or a rules based allocator with exposure and turnover controls. Second, language models can assist with constraint elicitation and documentation by translating investment beliefs and policy rules into machine readable constraints. Studies that evaluate the impact of conversational assistance on portfolio choices suggest that language models can improve portfolio hygiene when paired with clear prompts and risk constraints \parencite{Ko2024ChatGPTPortfolio}. In production settings the optimizer and the signal engine should remain distinct services with independent monitoring and fallback policies.

\subsection{Evaluation protocols and guardrails for live trading}
Trading evaluation must be time safe and economically grounded. Walk forward backtests should enforce document availability and embargo periods and they should report returns with explicit cost and impact assumptions, Sharpe and drawdown, turnover and capacity, and sensitivity to universe and rebalancing cadence. Work on lookahead bias in pretrained models and on time machine effects underscores the need for dated corpora and strict filters during both training and evaluation \parencite{Sarkar2024Lookahead,Drinkall2024TimeMachineGPT}. Cost and latency management are essential for live use and hybrid query routing can reduce spend while maintaining quality by steering easy queries to lightweight models and reserving high capacity models for hard cases \parencite{Ding2024HybridLLM}. Safety and governance require agent constitutions and risk aware judges that flag unsafe tool uses or policy violations \parencite{Hua2024TrustAgent,Yuan2024RJudge}. Systems should also detect contradictions in reports and maintain audit logs to support regulatory reviews \parencite{Deusser2023Contradictions,CaoFeinstein2024RegInterpret}. Strong baselines such as trend models help contextualize the incremental value of language driven workflows \parencite{Jiang2023PriceTrends}.

\subsection{Design patterns and practical guidance}
A robust design separates research and execution and binds language to verifiable evidence. Retrieval first prompting, tool verified numerics, and debate or critique before simulation reduce false positives. Confidence gating, materiality thresholds, and exposure caps stabilize portfolios. Human review remains important for new strategies, high impact actions, and regime changes. Regular stress tests and post trade analysis complete the loop and help teams decide when to promote a research signal into a production strategy.


\begin{center}
\begin{longtable}{@{}p{4.0cm}p{3.2cm}p{7.5cm}@{}}
\caption{Representative papers for LLM assisted trading and portfolio construction with one sentence summaries}
\label{tab:trading_papers_long} \\
\toprule
Paper & Contribution or setting & One sentence summary \\
\midrule
\endfirsthead

\multicolumn{3}{l}{\textit{Table \thetable\ continued}}\\
\toprule
Paper & Contribution or setting & One sentence summary \\
\midrule
\endhead

\midrule
\multicolumn{3}{r}{\textit{Continued on next page}}\\
\endfoot

\bottomrule
\endlastfoot

\textcite{Li2023TradingGPT} & Multi agent research to trade & A layered memory and role based framework proposes, critiques, and verifies trade ideas before execution in a controlled simulator. \\
\midrule
\textcite{Zhang2024FinAgent} & Tool augmented multimodal agent & A generalist agent integrates text and visuals and coordinates broker like tools under governance to produce executable decisions. \\
\midrule
\textcite{Yu2024FinMem} & Memory design for trading agents & A layered memory with character design improves persistence of analyst intent and boosts performance across multi day workflows. \\
\midrule
\textcite{Wang2024QuantAgent} & Self improving agent loop & A system that critiques prompts and strategies and that retests within a simulator yields more stable trading policies. \\
\midrule
\textcite{Wang2023AlphaGPT} & Human AI alpha mining & An interactive workflow uses critique and ranking to surface promising alphas with guardrails against overfitting. \\
\midrule
\textcite{Yuan2024AlphaGPT2} & Human in the loop alpha mining & The second version formalizes review gates and improves reliability when promoting ideas to production. \\
\midrule
\textcite{Liu2024StockAgent} & Trading in realistic environments & A benchmarked environment evaluates LLM based traders with market frictions and supports ablation studies. \\
\midrule
\textcite{Noguer2024CodeGenTrading} & Code generation for strategies & An empirical study shows that language models can scaffold trading code that passes unit tests when supervised by practitioners. \\
\midrule
\textcite{Yepes2024RAGChunk} & Retrieval aware analysis & A chunking method improves retrieval and long document analysis for filings and earnings research that feeds trading. \\
\midrule
\textcite{Wang2024DocLLM} & Layout aware modeling & A layout aware generator improves numerical reasoning over tables which reduces errors in research that precedes trades. \\
\midrule
\textcite{Nagy2023GenAILOB} & Limit order book simulation & A token level generative model of message flow produces realistic microstructure that is useful for execution policy testing. \\
\midrule
\textcite{Ding2024HybridLLM} & Cost and latency control & A hybrid routing approach reduces inference cost while maintaining answer quality which benefits live trading systems. \\
\midrule
\textcite{Hua2024TrustAgent} & Safety for agent systems & A constitution guided method constrains tool use and reduces unsafe actions during autonomous or semi autonomous operation. \\
\midrule
\textcite{Yuan2024RJudge} & Risk aware judging for agents & A benchmark and judge detect unsafe patterns in agent traces which complements trading evaluation. \\
\midrule
\textcite{Deusser2023Contradictions} & Governance and auditing & A contradiction detection pipeline highlights inconsistencies in financial reports and contributes to audit readiness. \\
\midrule
\textcite{CaoFeinstein2024RegInterpret} & Regulatory interpretation & A study outlines how language models can support interpretation of financial regulation which aids deployment governance. \\
\midrule
\textcite{Jiang2023PriceTrends} & Baseline for execution value add & A comprehensive trend study provides strong baselines that help measure the incremental value of language driven trading. \\
\end{longtable}
\end{center}

\section{Benchmarks and datasets for prediction and trading}
\label{sec:benchmarks}

The growth of financial language models has outpaced the availability of standardized and time safe benchmarks that connect textual understanding to tradable decisions. We organize the landscape into prediction oriented reasoning benchmarks that produce signals, trading and agent benchmarks that evaluate decision quality under constraints, and corpora and datasets that supply supervision or retrieval evidence. Across categories three design principles are foundational. First, temporal integrity ensures point in time documents and rolling and non overlapping out of sample evaluation with embargoed validation. Second, economically grounded metrics require profit and loss with costs, Sharpe, drawdown, turnover and capacity, and hit rate at realistic rebalancing frequencies. Third, reproducibility demands seeded data releases, fixed symbol universes with survivorship bias controls, and code to reconstruct splits.

\subsection{Prediction oriented reasoning and understanding}
A first class of resources evaluates whether models can extract and reason over financial information that plausibly feeds return prediction. FinQA targets numerical reasoning over text and tables and signals derived from correct KPI computation are often used upstream of event driven strategies \parencite{Chen2021FinQA}. FinanceBench and BizBench probe quantitative reasoning and business logic and they stress mathematical consistency that underlies valuation or surprise based signals \parencite{Islam2023FinanceBench,Koncel2023BizBench}. DocMathEval isolates long document numerical reasoning with tables which is a frequent failure point in earnings analysis \parencite{Zhao2023DocMathEval}. EconLogicQA evaluates economics sequential reasoning that matters for macro sensitive trade selection \parencite{Quan2024EconLogicQA}. The FinBen proposes a holistic financial benchmark that covers multiple tasks in finance \parencite{Xie2024TheFinBen}. AlphaFin frames analysis as a retrieval augmented stock chain that aligns evaluation with multi step reasoning workflows \parencite{Li2024AlphaFin}. These resources are not trading simulators and they measure signal fidelity since a failure on numerical reasoning or economic logic makes the trade premise unsound.

\subsection{Trading and agent evaluations}
Benchmarks that are tailored to trading decisions remain emergent. Agent frameworks report simulation results using internal market environments together with layered memory and tool use such as retrieval, backtesting, and data application programming interfaces \parencite{Li2023TradingGPT,Zhang2024FinAgent,Yu2024FinMem,Wang2024QuantAgent,Wang2023AlphaGPT,Yuan2024AlphaGPT2}. These works advance methodology through role specialization, verifier checks, and reflection and two gaps persist. First, there is limited standardization of market microstructure such as latency, slippage, queue priority, and the limit or market order mix. Second, there is heterogeneous choice of universes and horizons that complicates cross paper comparisons. Safety risk awareness for agents is emerging through R Judge which can complement trading evaluations by detecting unsafe tool usage or risk insensitive actions \parencite{Yuan2024RJudge}.

\subsection{Domain corpora and supervision for predictive pipelines}
Upstream datasets support sentiment, information extraction, event detection, and summarization that feed predictive engines. FiNER, FinRED, and REFinD supervise extraction of entities, relations, and events that populate knowledge graphs and enable cleaner point in time factors \parencite{Shah2023FiNERDataset,Sharma2022FinRED,Kaur2023REFinD}. ECTSum and MultiLing FNS provide summarization targets for earnings calls and reports, while MAEC and MONOPOLY supply multimodal earnings and policy material \parencite{Mukherjee2022ECTSum,ElHaj2019MultiLing,Li2020MAEC,Mathur2022MONOPOLY}. FinSBD focuses on structural boundary detection in unstructured filings and DocLLM demonstrates layout aware modeling that improves numerical question answering and KPI retrieval \parencite{Au2021FinSBD,Wang2024DocLLM}. These resources help construct evidence grounded signals that can survive audit.

\subsection{Multilingual and regional benchmarks}
Financial markets are multilingual and regulatory regimes differ across regions. Several efforts broaden coverage to non English disclosures. CFBenchmark, FinEval, and CFLUE provide Chinese financial evaluation resources and SuperCLUE Fin offers a fine grained analysis of Chinese tasks \parencite{Lei2023CFBenchmark,Zhang2023FinEval,Zhu2024CFLUE,Xu2024SuperCLUEFin}. Hirano constructs a Japanese financial benchmark that expands regional testing \parencite{Hirano2024JPBenchmark}. A study on bilingual prowess examines English and Spanish which is valuable for cross listings and American depositary receipts \parencite{Zhang2024Dolares}.

\subsection{Evaluation desiderata and a practical proposal}
A prediction and trading benchmark should enforce time safe document availability, include standardized universes, rebalancing schedules, and cost models, and report both signal metrics and portfolio metrics with ablations for retrieval, verifiers, and tool latency. It should publish agent traces with evidence links for auditability and include stress periods and regime slices together with multilingual tracks. A practical path is to couple AlphaFin or The FinBen style reasoning tasks with an open microstructure simulator and R Judge style safety checks.


\begin{center}
\begin{longtable}{@{}p{6.4cm}p{2.2cm}p{2.8cm}p{4.8cm}@{}}
\caption{Datasets and benchmarks that are most relevant to prediction and trading}
\label{tab:benchmarks_long} \\
\toprule
Resource & Modality & Primary task & Relevance to trading \\
\midrule
\endfirsthead

\multicolumn{4}{l}{\textit{Table \thetable\ continued}}\\
\toprule
Resource & Modality & Primary task & Relevance to trading \\
\midrule
\endhead

\midrule
\multicolumn{4}{r}{\textit{Continued on next page}}\\
\endfoot

\bottomrule
\endlastfoot

FinQA \parencite{Chen2021FinQA} & Text and tables & Numerical question answering & KPI correctness supports earnings surprise and event driven signals \\
\midrule
FinanceBench \parencite{Islam2023FinanceBench} & Text and numbers & Financial question answering & Valuation and logic checks help thesis validation \\
\midrule
BizBench \parencite{Koncel2023BizBench} & Text and numbers & Quantitative reasoning & Business logic consistency matters for fundamental theses \\
\midrule
DocMathEval \parencite{Zhao2023DocMathEval} & Long documents and tables & Numerical reasoning & Reduces miscalculation risk in filings driven research \\
\midrule
EconLogicQA \parencite{Quan2024EconLogicQA} & Text & Economics sequential reasoning & Supports macro sensitive selection and hedging decisions \\
\midrule
The FinBen \parencite{Xie2024TheFinBen} & Multi task & Holistic finance evaluation & Broad coverage aligns with diverse production workflows \\
\midrule
AlphaFin \parencite{Li2024AlphaFin} & RAG with stock chain & Financial analysis & Multi step reasoning for equity research with RAG \\
\midrule
FiNER and FinRED and REFinD \parencite{Shah2023FiNERDataset,Sharma2022FinRED,Kaur2023REFinD} & Text & IE and NER and relation extraction & Populates knowledge graphs and supports time safe factors and retrieval \\
\midrule
ECTSum and MultiLing FNS \parencite{Mukherjee2022ECTSum,ElHaj2019MultiLing} & Text & Summarization & Generates research briefs that accelerate analysis before trading \\
\midrule
MAEC and MONOPOLY \parencite{Li2020MAEC,Mathur2022MONOPOLY} & Audio and video and text & Multimodal earnings and policy & Supplies prosodic and policy cues for selection and sizing \\
\midrule
FinSBD and DocLLM \parencite{Au2021FinSBD,Wang2024DocLLM} & Text and layout & Structure detection and layout aware modeling & Stabilizes retrieval and improves numerical accuracy in long documents \\
\midrule
CFBenchmark and FinEval and CFLUE and SuperCLUE Fin and JP benchmark \parencite{Lei2023CFBenchmark,Zhang2023FinEval,Zhu2024CFLUE,Xu2024SuperCLUEFin,Hirano2024JPBenchmark} & Text & Regional evaluation & Enables non English disclosures and cross market strategies \\
\midrule
R Judge \parencite{Yuan2024RJudge} & Agent traces & Safety risk awareness & Adds guardrails for tool using agents during evaluation \\
\midrule
TradingGPT and FinAgent and FinMem and QuantAgent and Alpha GPT \parencite{Li2023TradingGPT,Zhang2024FinAgent,Yu2024FinMem,Wang2024QuantAgent,Wang2023AlphaGPT,Yuan2024AlphaGPT2} & Agent frameworks & Trading simulations & Provide methodology and protocols without standard data releases \\
\end{longtable}
\end{center}

\section{Challenges and open problems in LLM based prediction and trading}
\label{sec:challenges}

\subsection{Temporal leakage and time machine effects}
Return prediction with general web pretraining risks look ahead leakage because models may memorize future facts and surface them during prompting. Recent critiques show that even without explicit future documents at inference time the latent knowledge can leak into answers \parencite{Sarkar2024Lookahead,Drinkall2024TimeMachineGPT}. Effective mitigation combines corpora with strict publication cutoffs per evaluation fold, training data that is filtered by crawl date and source type, embargo windows for validation, and rationales that cite evidence published before the decision timestamp.

\subsection{Evaluation realism and economic significance}
Many studies report accuracy or correlation without a trading grade evaluation. Credible claims for language model signals require rolling walk forward backtests, conservative cost and impact models, turnover and capacity analysis, stress tests across regimes, and risk controlled performance with Sharpe, Sortino, drawdown, and tail loss. Benchmarks should include materiality filters so that statistically significant and economically trivial effects are not over interpreted.

\subsection{Faithfulness, hallucination, and numerical robustness}
Language models can produce confident but wrong rationales and can show brittle numerical reasoning. Evidence bound generation with citations to retrieved passages and tables, constrained tool use such as calculators and parsers, post hoc verification methods, and dual model cross checking reduce risk \parencite{Krishna2024GenAudit}. For trading the system should never change risk based on unverifiable rationales.

\subsection{Data coverage, point in time structure, and retrieval}
Filings, press releases, calls, and macro statements have heterogeneous formats and chunking and indexing must be point in time and stable across refactors. Layout aware encoders improve KPI extraction, structure boundary detection stabilizes retrieval, and financial information extraction datasets support higher precision evidence graphs \parencite{Wang2024DocLLM,Au2021FinSBD,Shah2023FiNERDataset,Sharma2022FinRED,Kaur2023REFinD}. Coverage gaps persist for small capitalization firms and non English issuers and multilingual resources help reduce these gaps.

\subsection{Cost, latency, and deployment economics}
Real time trading requires bounded latency and cost. Hybrid query routing can steer easy queries to cheaper models and reserve high capacity models for hard cases and low rank and quantized adaptation can further lower the footprint \parencite{Ding2024HybridLLM,Hu2021LoRA,Dettmers2024QLoRA,Ma2024OneBitLLMs}. System level reporting should include wall clock latency per decision and amortized compute cost per basis point of excess return.

\subsection{Interpretability, governance, and regulatory alignment}
Trading decisions must be explainable to risk, audit, and regulators. Desirable properties include rationales that are grounded in timestamped evidence, decomposition of effect that links evidence to predicted return, clear separation between signal generation and portfolio allocation, and audit logs for prompts, retrieved passages, and tool calls. Studies on regulatory interpretation and auditing support illustrate patterns for compliance ready pipelines \parencite{CaoFeinstein2024RegInterpret,Berger2023AuditLLM,Deusser2023Contradictions}.

\subsection{Security, privacy, and safety}
Financial language models raise attack surfaces that include prompt injection and alignment breaking attacks and they create privacy concerns. Agent frameworks need constitutions and safety checks to prevent unauthorized orders, personal data leakage, or policy violations \parencite{Hua2024TrustAgent,Yao2024LLMSecurity}. Ethical codes and evolving artificial intelligence regulations should inform deployment gates and operational controls.

\subsection{Robust generalization and regime shifts}
Language model signals can overfit a disclosure style, sector, or macro regime. Techniques that help include domain adaptation with retrieval from diverse sources, regime aware training through explicit slicing or adversarial invariance, multilingual modeling for cross listed firms, and ensembling with classical factors to stabilize exposures. Reporting should include sector breakdowns and regime wise performance.

\subsection{Data synthesis and augmentation}
Language model based augmentation can improve label efficiency and synthetic data can introduce biases or leakage if generated with non time safe context. Synthetic examples should be marked, confined to training, and stress tested for bias. Evaluation sets should never contain synthetic items.


\subsection{Minimum reporting standard}
As a minimum reporting standard, studies should enforce time safe data and splits with document availability at the decision timestamp; present a full cost model with commissions, spreads, and market impact calibrated to the universe and size; report turnover, capacity, and the effect of transaction costs on net performance; analyze stress periods and regimes with sector level breakdowns; provide evidence grounded rationales and verified calculations for key examples; include ablations for retrieval, verifiers, and query routing together with wall clock latency and compute cost per decision; release seeds and code to reconstruct time splits and point in time indices or a protocol that supports replication; and compare against strong trend and factor baselines to quantify incremental value.

\section{Conclusion}
\label{sec:conclusion}

Large Language Models are redefining quantitative investing by turning unstructured financial information into auditable signals and coordinated actions. In the new quant, language models do not replace classical statistics or portfolio theory and they compose with them. Models read filings and calls, cite timestamped evidence, invoke calculators and parsers for numerics, and hand verified signals to risk aware allocation engines. Evidence accumulated in recent work suggests real potential for excess returns in selected regimes and universes, especially when models are domain adapted, retrieval grounded, and evaluated with trading grade procedures.

The field should adopt three practical principles. Separate concerns means keeping signal generation with retrieval and verification distinct from portfolio construction so that objectives and accountability remain clear. Bind language to evidence means requiring timestamped citations and tool verified calculations before any position changes and logging prompts, retrievals, and tool calls for auditability. Evaluate like a practitioner means enforcing time safe splits with document availability checks and realistic costs and slippage, reporting turnover and capacity, analyzing stress regimes, and disclosing latency and compute cost per decision rather than only reporting accuracy.

A focused research program follows from these principles. The community should design standardized prediction to trading benchmarks that couple reasoning tasks such as FinQA, FinanceBench, and AlphaFin with open and time safe market simulators and with safety audits that detect risky tool use \parencite{Chen2021FinQA,Islam2023FinanceBench,Li2024AlphaFin,Yuan2024RJudge}. Training and evaluation should emphasize temporal robustness through filtered corpora, explicit publication cutoffs, and diagnostics for look ahead effects \parencite{Sarkar2024Lookahead,Drinkall2024TimeMachineGPT}. Explainable financial language models should produce evidence anchored rationales that map to portfolio exposures to meet governance needs. Multilingual and low resource finance should receive sustained attention to support global coverage and cross listing dynamics. Systems should be cost aware through hybrid query routing and efficient adaptation methods such as low rank tuning, quantization aware finetuning, and one bit optimization \parencite{Ding2024HybridLLM,Hu2021LoRA,Dettmers2024QLoRA,Ma2024OneBitLLMs}. Human and AI collaboration should be central and analyst in the loop critique and debate agents can increase faithfulness without sacrificing speed while agent constitutions and judges improve safety \parencite{Hua2024TrustAgent,Yuan2024RJudge}.

With transparent benchmarks, temporal discipline, and audit ready system design, financial language models can progress from promising prototypes to reliable building blocks in modern investment processes. The promise of the new quant is to translate textual understanding into robust, risk controlled, and economically meaningful trades.

\section*{Disclosure}
Portions of this paper were drafted or paraphrased with the assistance of ChatGPT (OpenAI). 
The author reviewed, edited, and takes full responsibility for the intellectual content and conclusions presented in this work.

\printbibliography

\end{document}